\documentclass[10pt]{iopart}
\usepackage{iopams} 
\usepackage{graphicx}
 
\begin{document}

\title[Fractional Spin ]{Fractional spin - a property of particles described with a fractional Schr\"odinger equation}

\author{Richard Herrmann}

\address{
GigaHedron, Farnweg 71, D-63225 Langen, Germany
}
\ead{herrmann@gigahedron.com}
\begin{abstract}
It is shown, that the requirement of invariance under spatial rotations reveales an intrinsic fractional 
extended translation-rotation-like property for particles described with the fractional Schr\"odinger equation, 
which we call 
fractional spin.
\end{abstract}

%

\section{Introduction}
The fractional calculus \cite{f3},\cite{he08} provides a set of axioms and methods to extend the coordinate and corresponding
derivative definitions in a reasonable way from integer order n to arbitrary order $\alpha$:
\begin{equation}
\{ x^n, {\partial^n \over \partial x^n} \} 
\rightarrow
\{ x^\alpha, {\partial^\alpha \over \partial x^\alpha} \}
\end{equation}
The definition of the fractional order derivative is not unique,  several definitions 
e.g. the Riemann, Caputo, Weyl, Riesz, Gr\"unwald  fractional 
derivative definition coexist \cite{f1}-\cite{grun}.
To keep this paper as general as possible,   we do not apply a 
specific representation of the fractional derivative operator.

We will only assume, that an appropriate mapping on real numbers of coordinates $x$ and fractional coordinates $x^\alpha$ 
and
functions $f$ and fractional derivatives $g$  exists  and
that a Leibniz product rule is defined properly.  Therefore we use $x^\alpha$ as a short hand notation for e.g.
$sign(x) |x|^\alpha$ as demonstrated in \cite{he07} and $\partial^\alpha _x=\partial^\alpha / \partial x^\alpha$ as a 
short hand notation for e.g. the fractional
left and right Liouville derivative ($D_{+}^\alpha,D_{-}^\alpha$):
\begin{eqnarray} 
\label{liouville1}
(D_{+}^\alpha f)(x) &=&  
\frac{1}{\Gamma(1 -\alpha)} \frac{\partial}{\partial x}  
     \int_{-\infty}^x  d\xi \, (x-\xi)^{-\alpha} f(\xi)\\
\label{liouville2}
(D_{-}^\alpha f)(x) &=&  
\frac{1}{\Gamma(1 -\alpha)} \frac{\partial}{\partial x}  
     \int_x^\infty  d\xi \, (\xi-x)^{-\alpha} f(\xi)
\end{eqnarray} 
which may be combined via
\begin{eqnarray}
\partial^\alpha_x f(x) &=& {D_+^\alpha - D_-^\alpha \over 2 \sin(\alpha \pi/2) } f(x) \\
&=&   \Gamma(1+\alpha) {\cos(\alpha \pi/2) \over \pi } \int_0^\infty {f(x+\xi) - f(x-\xi)  \over \xi^{\alpha+1}} d\xi \\
&& \qquad \qquad\qquad \qquad\qquad \qquad \qquad  0\leq \alpha < 1 \nonumber
\end{eqnarray}
For this derivative definition, the invariance of the scalar product follows:
\begin{equation}
\label{scalar}
\int_{-\infty}^{\infty}
\left( {\partial^\alpha \over \partial x^\alpha}^*  f^{*}(x) \right) g(x) dx =
-\int_{-\infty}^{\infty}
f(x)^{*} \left( {\partial^\alpha \over \partial x^\alpha}  g(x) \right)  dx 
\end{equation}
where  $^{*}$ denotes the complex conjugate. 

The Leibniz product rule is used in the following form \cite{f3},\cite{o1}:
\begin{equation}
\label{leibniz}
{\partial^\alpha \over \partial x^\alpha} (\phi \psi) = \sum_{k=0}^{\infty} 
\left(\begin{array}{c}
\alpha \\ k \\
\end{array} \right)
( {\partial^k \over \partial x^k} \phi ) 
( {\partial^{\alpha-k} \over \partial x^{\alpha-k}}\psi ) 
\end{equation}
where the fractional binomial is given by
\begin{equation}
\left(\begin{array}{c}
\alpha \\ k \\
\end{array} \right)
= {\Gamma(1+\alpha) \over \Gamma(1+k) \Gamma(1+\alpha-k)}
\end{equation}
and $\Gamma(z)$ is the Euler $\Gamma$-function.

We define the following set of conjugated operators on an euclidean space for $N$ particles in space coordinate
representation:
\begin{eqnarray}
\hat{P}_\mu &=& \{  \hat{P}_0,\hat{P}_i\} = \{ i \hbar \partial_t,-i \left( \frac{\hbar}{m c} \right)^{\alpha} m c  \, \partial^\alpha_i \}\\
\hat{X}_\mu &=& \{  \hat{X}_0,\hat{X}_i\} = \{ t, \left( \frac{\hbar}{m c} \right)^{(1-\alpha)} (x_i^\alpha)  \} \\
& & \qquad\qquad\qquad\qquad\qquad i = 1,..., 3 N \nonumber
\end{eqnarray}
Due to (\ref{scalar}), these operators are hermitean.

With these operators, the classical, non relativistic Hamilton function $H_c$, which depends on
the classical momenta and coordinates $\{p_i,x^i\}$
\begin{equation}
H_c = \sum_{i=1}^{3N}\frac{p^2_i}{2m} +V(x^1,...,x^i,...,x^{3N}) 
\end{equation}
is quantized. This yields the Hamiltonian $H^\alpha$ 
\begin{equation}
H^\alpha = -\frac{1}{2} m c^2 \left( \frac{\hbar}{m c} \right)^{2 \alpha} 
\sum_{i=1}^{3N}
\partial^\alpha_i \partial^\alpha_i +V(\hat{X}^1,...,\hat{X}^i,...,\hat{X}^{3N})
\end{equation}
Thus, a time dependent Schr\"odinger type equation for fractional derivative 
operators results
\begin{equation}
\fl
\label{sgl}
H^\alpha \Psi = (-\frac{1}{2} m c^2 \left( \frac{\hbar}{m c} \right)^{2 \alpha} 
\sum_{i=1}^{3N}
\partial^\alpha_i \partial^\alpha_i  +V(\hat{X}^1,...,\hat{X}^i,...,\hat{X}^{3N})) \Psi = i \hbar \partial_t \Psi 
\end{equation}
For $\alpha=1$ this reduces to the classical Schr\"odinger equation.

\section{The internal structure of fractional particles}
Properties of particles, which are described by wave equations, may be investigated using the
commutation relations of fundamental symmetry operations.

Lets call a particle elementary, if it is described by a potential- and field-free wave equation. If in addition
there is an internal structure, which is determined by additional quantum numbers, it may be revealed
e.g. considering the behaviour under rotations.

We will investigate the most simple case, the behaviour of the fractional free Schr\"odinger equation (\ref{sgl})
for a single particle ($V=0$, $N=1$) under rotations in $R^2$ about the z-axis.

Using the Leibniz product rule (\ref{leibniz}) the fractional derivative of the
product $x f(x)$ is given by 
\begin{eqnarray}
\label{leibnizxf}
\partial^\alpha_x (x f(x)) &=& \sum_{j=0}^\infty    
\left(\begin{array}{c}
\alpha \\ j \\
\end{array} \right)
(\partial^j_x x) \partial^{\alpha-j}_x f(x) \\
&=& \left(x \partial^\alpha_x + \alpha \partial^{\alpha-1}_x \right) f(x)
\end{eqnarray}
We define a generalized fractional angular momentum operator 
$K^\beta$ with z-component $K^\beta_z$
\begin{equation}
\label{kz}
K^\beta_z = i \left({\hbar \over mc}\right)^\beta \, mc\,  (x \partial^\beta_y - y \partial^\beta_x) 
\end{equation}
The compoments $K^\beta_x, K^\beta_y$ are given by cyclic permutation of the spatial indices in (\ref{kz}). 
Using (\ref{leibnizxf})
the commutation relation  with the Hamiltonian $H^\alpha$ of the free Schr\"odinger equation (\ref{sgl}) results as
(with units $m=1$, $\hbar=1$, $c=1$):
\begin{eqnarray}
\label{kkk}
[-i 2 K^\beta_z, H^\alpha] &=& [K^\beta_z, \partial^{2 \alpha}_x + \partial^{2 \alpha}_y + \partial^{2 \alpha}_z ]\nonumber\\
                      &=& [K^\beta_z, \partial^{2 \alpha}_x + \partial^{2 \alpha}_y]\nonumber\\
                      &=& K^\beta_z (\partial^{2 \alpha}_x + \partial^{2 \alpha}_y)
                      - (\partial^{2 \alpha}_x + \partial^{2 \alpha}_y)K^\beta_z \nonumber\\
                      &=& (x \partial^\beta_y - y \partial^\beta_x) (\partial^{2 \alpha}_x + \partial^{2 \alpha}_y)
                           - (\partial^{2 \alpha}_x + \partial^{2 \alpha}_y)(x \partial^\beta_y - y \partial^\beta_x)\nonumber\\
                      &=& x \partial^{ 2 \alpha}_x \partial^\beta_y
                        + x \partial^{ 2 \alpha+ \beta}_y 
                        - y \partial^{ 2 \alpha+ \beta}_x 
                        - y \partial^\beta_x \partial^{ 2 \alpha}_y \nonumber\\ 
                      & &  - \left(    
                          \partial^{2 \alpha}_x x  \partial^\beta_y 
                         -  y \partial^{ 2 \alpha+ \beta}_x 
                         +  x \partial^{ 2 \alpha+\beta}_y 
                         - \partial^{2 \alpha}_y y \partial^\beta_x
                          \right) \nonumber \\ 
                      &=& x \partial^{ 2 \alpha}_x \partial^\beta_y
                        - y \partial^\beta_x \partial^{ 2 \alpha}_y \nonumber\\ 
                      & &  - \left(    
                          x \partial^{2 \alpha}_x   \partial^\beta_y 
                          + 2 \alpha  \partial^{2 \alpha-1}_x   \partial^\beta_y 
                         - y \partial^\beta_x \partial^{2 \alpha}_y 
                         - 2 \alpha \partial^\beta_x \partial^{2 \alpha-1}_y 
                          \right) \nonumber \\
                      & = &  - 2 \alpha \left(    
                          \partial^{2 \alpha-1}_x   \partial^\beta_y 
                         - \partial^\beta_x \partial^{2 \alpha-1}_y 
                          \right) 
 \end{eqnarray}
Setting $\beta=1$ we obtain for the z-component of the standard angular momentum operator  $L_z$ the 
commutation relation
\begin{equation}
[L_z,H^\alpha] = [K^{\beta=1}_z,H^\alpha] =  -  i \alpha \left(    
                          \partial^{2 \alpha-1}_x   \partial_y 
                         - \partial_x \partial^{2 \alpha-1}_y 
                          \right) 
\end{equation}
which obviously is not vanishing. Therefore particles described with the 
fractional Schr\"o\-dinger equation (\ref{sgl}) contain an internal structure for $\alpha \neq 1$.

We now define the fractional total angular momentum 
$J^\beta$ 
with z-component $J^\beta_z$.
Setting 
\begin{equation}
\beta = 2 \alpha-1
\end{equation}
we obtain with $J^{2 \alpha-1}_z = K^{2 \alpha-1}_z$, 
and with (\ref{kkk}) 
\begin{equation}
[J^{2 \alpha-1}_z,H^\alpha ] = 0
\end{equation}
this operator commutes with the Hamilton operator. Therefore $J^{2 \alpha-1}_z$ indeed is the
fractional analogue of the standard z-component of the total angular momentum.

Now we define the z-component of a fractional intrinsic angular momentum $S^\beta_x$
with
\begin{equation}
J^{2 \alpha-1}_z = L_z + S^{2 \alpha-1}_z 
\end{equation}
The explicit form is given by
\begin{eqnarray}
S^{2 \alpha-1}_z &=&  i 
 x \left(\left({\hbar \over mc}\right)^{2 \alpha-1} \! mc\, \partial^{2 \alpha-1}_y - \hbar \partial_y\right) \nonumber \\
&-&
i y \left(\left({\hbar \over mc}\right)^{2 \alpha-1} \! mc\,\partial^{2 \alpha-1}_x - \hbar \partial_x\right)   
\end{eqnarray}
This operator vanishes for $\alpha=1$, whereas for $\alpha \neq 1$ it gives the z-component of a 
fractional spin.

Lets call the difference between fractional and ordinary derivative  $\delta p$, or more precisely
the components
\begin{equation}
\delta p_i = i  \left(\left({\hbar \over mc}\right)^{2 \alpha-1} mc\,\partial^{2 \alpha-1}_i - \hbar \partial_i \right) 
\end{equation}
for $S^{2 \alpha-1}_z$ we can write
\begin{equation}
S^{2 \alpha-1}_z =   x \delta p_y - y \delta p_x 
\end{equation}
The components of a fractional spin vector are then given by the cross product
\begin{equation}
\vec{S}^{2 \alpha-1} =   \vec{r} \times \delta \vec{p}
\end{equation}
Therefore fractional spin describes an internal fractional rotation, which is proportional to the 
momentum difference between fractional and ordinary  momentum. For a given $\alpha$ it has exactly
one component. 

With $J^{2 \alpha-1}_x= K^{2 \alpha-1}_x$ and $J^{2 \alpha-1}_y=K^{2 \alpha-1}_y$, 
the commutation relations for the total fractional angular momentum are given by
\begin{eqnarray}
\left[ J^{2 \alpha-1}_x, J^{2 \alpha-1}_y \right] &=&  (2 \alpha-1) {\hbar \over m c} J^{2 \alpha-1}_z p_z^{2(\alpha-1)} \\
\left[ J^{2 \alpha-1}_y, J^{2 \alpha-1}_z \right] &=&  (2 \alpha-1) {\hbar \over m c} J^{2 \alpha-1}_x p_x^{2(\alpha-1)} \\
\left[ J^{2 \alpha-1}_z, J^{2 \alpha-1}_x \right] &=&  (2 \alpha-1) {\hbar \over m c} J^{2 \alpha-1}_y p_y^{2(\alpha-1)} 
\end{eqnarray}
with components of the momentum operator $p$ or generators of fractional translations respectively given as
\begin{equation}
p_i^\beta =   i \left( {\hbar \over m c} \right)^\beta m c \, \partial^\beta_i 
\end{equation}
Therefore in the general case $\alpha \neq 1$ an extended fractional rotation group is generated, which contains 
an additional fractional translation factor.

\section{Conclusion}
We conclude, that fractional elementary particles which are decribed with the fractional Schr\"odinger equation, 
carry an internal structure, which we call
fractional spin, because analogies to e.g. electron spin are close. 

Consequently, the transformation properties of the fractional Schr\"o\-di\-nger equation 
are more related to the ordinary Pauli-equation
than to the ordinary ($\alpha=1$) Schr\"o\-dinger-equation.  


\end{document}